\documentclass{sig-alt-gov2}
\usepackage{times}
\usepackage[latin1]{inputenc}
\usepackage{amsmath}
\usepackage{amsfonts}
\usepackage{amssymb}
\usepackage{verbatim}
\usepackage{algorithm}
\usepackage{algpseudocode}
\usepackage{graphicx}
\usepackage{hyperref}
\usepackage{subfigure}
\usepackage{url}
\usepackage{epsfig}
\usepackage{cite}
\algnotext{EndFor}
\algnotext{EndIf}
\algnotext{EndWhile}
\usepackage{pdfpages}

\begin{document}

\conferenceinfo{DyNetMM'13,} {June 23, 2013, New York, USA.} 
\CopyrightYear{2013} 
\crdata{978-1-4503-2037-5/13/06} 
\clubpenalty=10000 
\widowpenalty = 10000

\title{Fast Search for Dynamic Multi-Relational Graphs}

\numberofauthors{4} 
\author{
%
%
\alignauthor
Sutanay Choudhury\\
       \affaddr{Pacific Northwest National Laboratory, USA}\\
       \email{sutanay.choudhury@pnnl.gov}
\alignauthor
Lawrence Holder\\
       \affaddr{Washington State University, USA}\\	
       \email{holder@wsu.edu}
\alignauthor
George Chin\\
       \affaddr{Pacific Northwest National Laboratory, USA}\\
       \email{george.chin@pnnl.gov}
\and  
\alignauthor
John Feo\\
       \affaddr{Pacific Northwest National Laboratory, USA}\\
       \email{john.feo@pnnl.gov}
}
\maketitle
\thispagestyle{empty}

\begin{abstract}
Acting on time-critical events by processing ever growing social media or news streams is a major technical challenge.  Many of these data sources can be modeled as multi-relational graphs.  Continuous queries or techniques to search for rare events that typically arise in monitoring applications have been studied extensively for relational databases. This work is dedicated to answer the question that emerges naturally: \textsl{how can we efficiently execute a continuous query on a dynamic graph?}  This paper presents an exact subgraph search algorithm that exploits the temporal characteristics of representative queries for online news or social media monitoring.  The algorithm is based on a novel data structure called the \emph{Subgraph Join Tree (SJ-Tree)} that leverages the structural and semantic characteristics of the underlying multi-relational graph.  The paper concludes with extensive experimentation on several real-world datasets that demonstrates the validity of this approach.
\end{abstract}


\section{Introduction}
Social networks, social media websites and mainstream news media are driving an exponential growth in online content. This information barrage presents both a formidable challenge and an opportunity to applications that thrive on situational awareness.  Examples of such applications include emergency response, cyber security, intelligence and finance where the data stream is monitored continuously for specific events.  Timeliness of the detection carries paramount importance for such applications.  The applications derive their competitive edge from fast detection as late detection may not have much value due to incurred damage to resources.  Our work is motivated by queries that look for rare events, have a time constraint on the time to discovery and never need a bulk retrieval of historic data due to their monitoring nature.  

\textit{Continuous queries} evolved in the field of relational databases to address applications with precisely the above characteristics.  A \textit{continuous query system} is defined as one where a query logically runs continuously over time as opposed to being executed intermittently \cite{Law:2011:RLD:1966385.1966386, Terry:1992:CQO:141484.130333}.  Thus, continuous query processing is data-driven or trigger oriented.  Many of the prominent news or social media streams can be represented as multi-relational data sources.  
Multi-relational graphs are often an attractive representation for data sources with sparsity.  The problem of monitoring events in such data streams can be viewed as continuously searching the dynamic graph for patterns that represent events of interest. 

\begin{figure}[]
\centering
\includegraphics[scale=0.45]{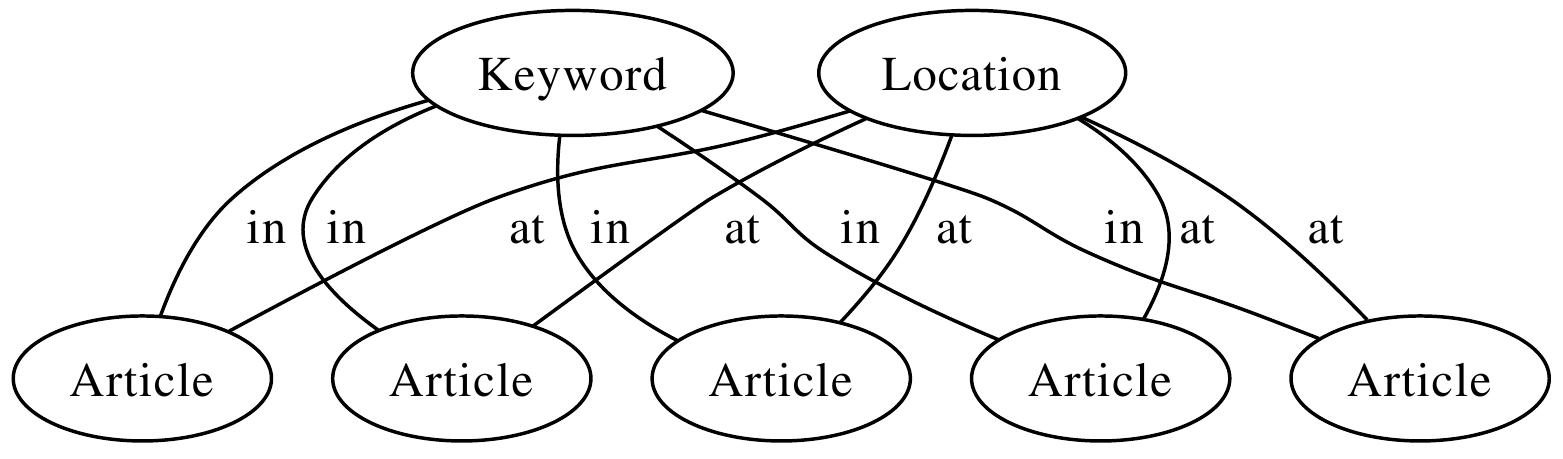}
\caption{A graph query for monitoring emergencies in social media and news streams.}
\label{fig:fig_2_query}
\end{figure}

Fig. \ref{fig:fig_2_query} shows a graph pattern that represents such an event. An operator may substitute the "keyword" with "fire" or "accident" and register several queries.  Articles refer to articles in news as well as social media posts.  This query will capture events that are reported from the same location.  Observe that we specify the label for only one vertex in this query, the rest of the vertices have only type specified.  Therefore, we are using the labeled vertex to anchor into a context and report when multiple events with that context are detected in the data stream.

\textit{Graph search} involves finding exact or approximate matches for a query subgraph in a larger graph.  It has been studied extensively and is formally defined as the problem of \textit{subgraph isomorphism}: given a  pattern or query graph (henceforth described as query graph) $G_q$ and a larger input graph (henceforth described as the data graph) $G_d$, find all isomorphisms of $G_q$ in $G_d$.  Following the definition of isomorphism, the matching involves finding a one-to-one correspondence between the vertices of a subgraph of $G_d$ and vertices of $G_q$ such that all vertex adjacencies are preserved.  Now consider the challenges in applying traditional graph search techniques to this problem.  Unless carefully adapted, a standard search function will search the entire data graph repeatedly and retrieve the same search results.  Also, many of the best performing graph search algorithms rely on indexing the graph.  Even with an interval as large as 5 minutes, rebuilding the index of a massive graph repeatedly is  infeasible.  This motivates exploration of incremental algorithms for continuous queries, although the general problem of incremental subgraph isomorphism is proven to be NP-complete as well \cite{Fan:2011:IGP:1989323.1989420}.

Queries like the one shown in Fig. \ref{fig:fig_2_query} share a number of common attributes.  First, they all involve an implicit time window to suggest the timeliness aspect associated with the query.  Clearly, the length of the time window varies depending on the application context.  The average monitoring time window for a high volume social media stream may be in tens of minutes whereas the equivalent period for online news may be in hours or days.  Second, all these queries aim to discover a number of temporal events that share the same context, such as a common set of keywords and location.  Lastly, a multi-relational graph often takes the form of k-partite graphs \cite{DBLP:conf/sigmod/ZhaoLXH11, DBLP:conf/icdm/SpyropoulouB11}  where each partite set represents a group of entities of the same type.  For queries as ones described in Fig. \ref{fig:fig_2_query}, each event that is represented by an article or a tweet can be viewed as a k-partitite subgraph.

We exploit these three features to implement a continuous query processing framework for multi-relational graphs.  First, by utilizing a rolling time window we continuously prune partial search results that would otherwise need to be tracked and would contribute to the combinatorial growth in memory utilization.  Second, the temporal property of the vertices and edges representing events suggests that it is logical to search for distinct subgraphs where such ``temporal" vertices or edges are ordered, thus significantly reducing the search space.  Finally, we take advantage of the multi-relational structure of the data and the characteristics of temporal events to avoid expensive joins.  Given a multi-relational query graph we decompose it in a hierarchical fashion.  We design a data structure called the \emph{Subgraph Join Tree}, or henceforth referred as the \emph{SJ-Tree} to model the hierarchical decomposition and store matches with various subgraphs of the query graph as represented in the tree.  This paper demonstrates the validity of this decomposition approach towards query processing.  Automated construction of SJ-Tree from an arbitrary query graph is not discussed in this paper.  We refer to the smallest units of the decomposed query graph as ``search primitives", which almost always consists of more than one edge.  As new edges arrive over time, we continuously perform (a) ``local searches" to look for matches with the search primitives and (b) use the decomposition structure to ``join" them into progressively larger matches.  This represents a middle ground between the periodic application of a graph search algorithm on the data graph and the approach that would have been employed by a traditional stream database.  Stream databases have no alternative but to model each edge in the query graph as a separate join operator.  Our model can express this degenerate case where an edge is represented as a search primitive in the SJ-Tree, but the performance is extremely poor.   By grouping subgraphs into search primitives, we can simplify the query plan, significantly improve performance by multiple orders of magnitude, and perhaps most importantly, reason about the trade-offs involved and explore a large space of possible optimizations.

\subsection{Contributions}
Our contributions from this research are summarized below.

1. We introduce a data structure called \textsl{SJ-Tree} for query graph decomposition (section \ref{sec:Incremental Query Processing}) and present a novel subgraph search algorithm (\ref{sec:Continuous Query Algorithm})  for continuous queries on dynamic multi-relational graphs.

2. We compare our performance with the incremental subgraph isomorphism algorithm developed by Fan et al. \cite{Fan:2011:IGP:1989323.1989420} and show that our approach provides improvements by multiple orders of magnitude (section \ref{sec:Experimental Results}).

3. We present a series of experiments on representative online news (New York Times), co-authorship networks (DBLP) and social media data sources (Tencent Weibo) modeled as multi-relational graphs.  The scale of these datasets are orders of magnitude larger than previously reported research \cite{Chen:2010:CSP:1850481.1850517, Fan:2011:IGP:1989323.1989420} in the literature (section \ref{sec:Experimental Results}).
	
4. We present a theoretical model for complexity analysis of the search algorithm.  We also provide an extensive experimental analysis of the algorithm's performance as a function of the frequency distribution of vertex labels for verification of the theoretical model.  

\begin{figure}[ht]
\centering
\subfigure{} {
\includegraphics[scale=0.4]{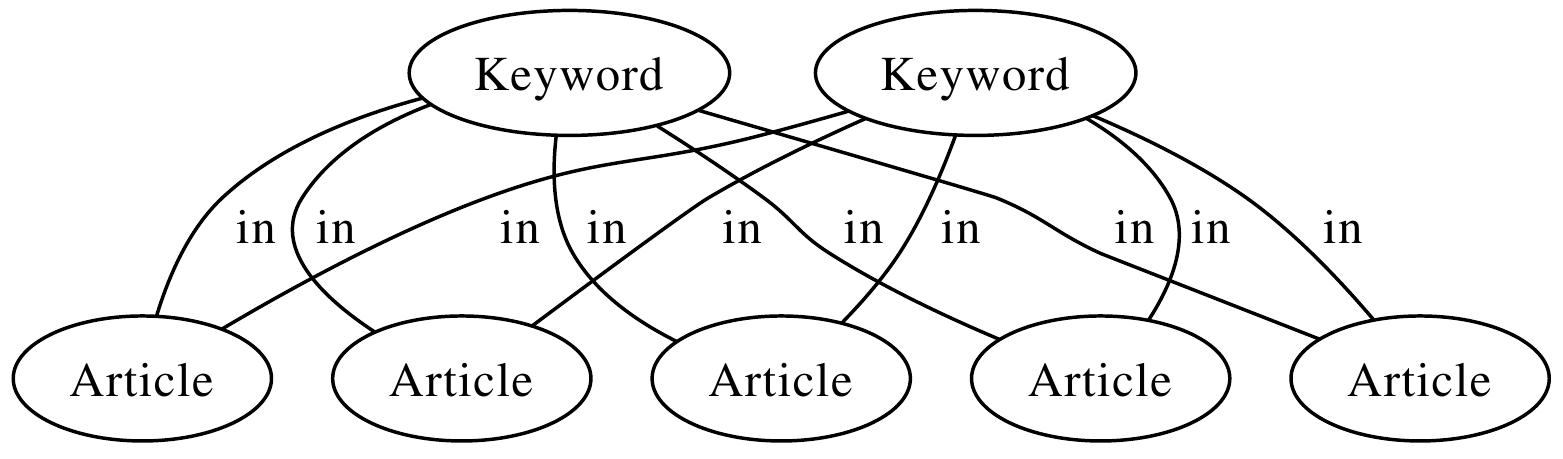}
\label{fig:subfig1}
}
\centering
\subfigure{} {
\includegraphics[scale=0.4]{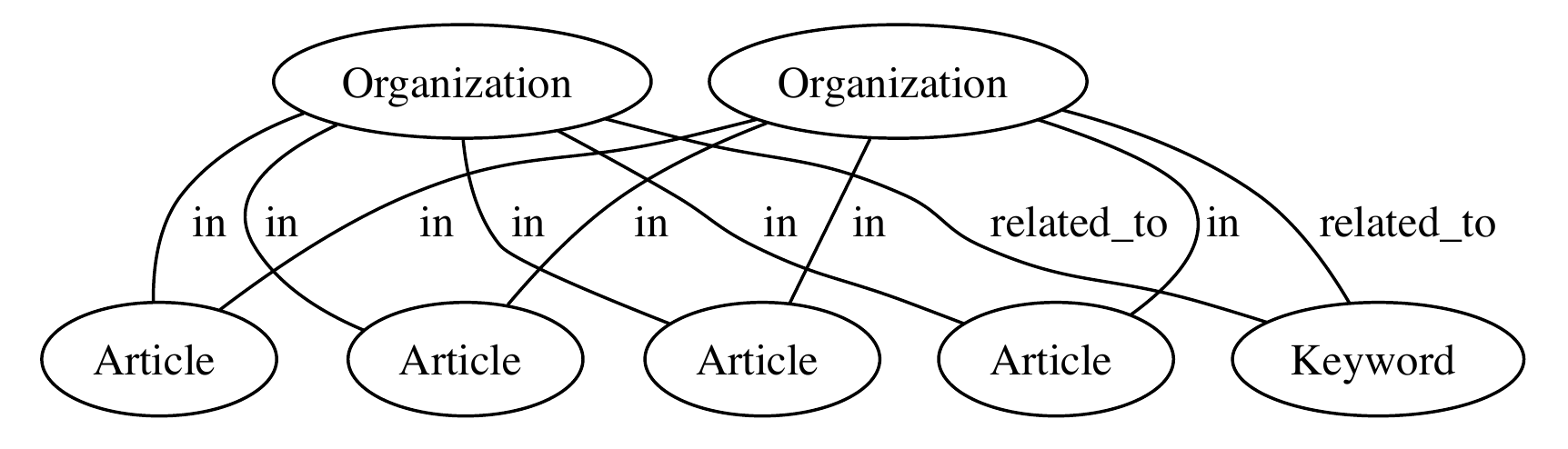}
\label{fig:subfig2}
}
\label{fig:example_queries}
\caption[]{More examples of monitoring queries on multi-relational graphs.  The query at the top can be used to discover events in a certain context.  Set one of the keywords to ``Oil" and run the query to discover various events that center around oil, such as price movements, discovery, accident etc.   By setting the keyword to ``buyout", the bottom query can be used to detect when news surface about a merger between two companies. }
\end{figure}

\subsection{Problem Statement}
\label{subsec:Problem Statement}

Every edge in a dynamic graph has a timestamp associated with it and therefore, for any subgraph $g$ of a dynamic graph we can define a time duration $\tau(g)$ which is equal to the duration between the earliest and latest edge belonging to $g$.  Given a dynamic multi-relational graph $G_d$, a query graph $G_q$ and a time window $t_W$, we report whenever a subgraph $g_d$ that is isomorphic to $G_q$ appears in $G_d$ such that $\tau(g_d) < t_W$.  The isomorphic subgraphs are also referred to as \textit{matches} in the subsequent discussions.  If $M(G^k_d)$ is the cumulative set of all matches discovered until time step $k$ and $E_{k+1}$ is the set of edges that arrive at time step $k+1$, we present an algorithm to compute a function $f\left(G_d, G_q, E_{k+1}\right)$ which returns the incremental set of matches that result from updating $G_d$ with $E_{k+1}$ and is equal to $M(G^{k+1}_d) - M(G^k_d)$.  We assume that the graph only receives edge inserts and no deletions.

\section{Background}
\label{sec:background}

\subsection{Multi-Relational Graphs}

Single relational graphs have been widely used to model systems comprised of homogeneous elements related by a single type of relation.  A social network where vertices represent people and edges represent connections between people is an example of a single-relational graph.  A multi-relational graph becomes a useful construct for modeling heterogeneous relations between a possibly heterogeneous set of entities. 

\textsc{Definition 2.1.1 } \textbf{Multi-Relational Graph} A multi-relational graph denoted as $G = (V, E)$, is a graph representation of a multi-relational database.  If the database contains $K$ entity types as $\xi_1, ...\xi_K$, then the vertex set $V(G)$ is partitioned into $K$ sets 
$V_1, ..., V_K$.  For any vertex $v \in V_k$, the label for the vertex is represented from the domain of the entity type $\xi_k$.  The edges of the graph are the relations between various entities as indicated in an entity-relation model.  Thus, an edge in the graph $e \in E(G), e = \left(v_i, v_j\right)$ is an instance of a relation $R_{ij}$ between entities $\xi_i$ and $\xi_j$. 

A graph representation of such a multi-relational database takes the form of a K-partite graph \cite{DBLP:conf/icdm/SpyropoulouB11}, if there are no relations between homogeneous entities or equivalently, if there are no edges between vertices that belong to the same partite set.  In practice, such relationships are not rare.  Examples of such linkages are citation links between articles and social ties between two members in a network.  However, we omit unary relationships from our multi-relational model.   Our omission of unary relationships is driven by usability and a desire for simplicity.  Fig. 1 and 2 show a number of examples embodying a range of events.  Consider the example in Fig. 2 that detects a series of articles that refer to the same set of keywords;  one may wish to introduce unary relationships in the graph to indicate citation between articles and thus, focus only on articles with high citation counts.  However, such queries can be alternatively represented by adding a query constraint to the vertices that require them to have a minimum degree.  Or, such relations could be represented using an intermediate vertex of a different type.  Thus, for the scope of this work we define patterns of interest as query graphs that are subgraphs of the K-partite multi-relational graph.  

\subsection{Continuous Queries}

A continuous query can be described as computing a function $f$ over a stream $S$ continuously over time and notifying the user whenever the output of $f$ satisfies a user-defined constraint \cite{Law:2011:RLD:1966385.1966386}.  They are distinguished from ad-hoc query processing by their high selectivity (looking for unique events) and need to detect newer updates of interest as opposed to retrieving lots of past information.  In this paradigm the primary objective is to notify a listener as soon as the query is matched.  One may view conventional databases as passive repositories with large collections of data that work in a request-response model whereas continuous queries are data-driven or trigger oriented.  These features coupled with real-time demands challenge many of the fundamental assumptions for conventional databases and establish continuous query processing on relational data streams as a major research area.   The literature on database research from the past two decades is abundant with work on continuous query systems \cite{Chandrasekaran:2003:TCD:872757.872857, DBLP:conf/sigmod/ArasuBBDIRW03}.  Babcock et al. \cite{Babcock:2002:MID:543613.543615} provide an excellent overview of continuous query systems and their design challenges. 

\subsection{Graph Queries}

Graph querying techniques have been studied extensively in the field of pattern recognition over nearly four decades \cite{conte2004thirty}.  Our work is focused on \textsl{subgraph isomorphism} which is as defined as follows.

\textsc{Definition 2.2.1 } \textbf{Subgraph Isomorphism} Given the query graph $G_q$ and a matching subgraph of the data graph ($G_d$) denoted as $G^{'}_d$, a matching between $G_q$ and $G^{'}_d$ involves finding a bijective function $f : V(G_q) \rightarrow V(G^{'}_d)$ such that for any two vertices $u_1, u_2 \in V(G_q)$,  $(u_1, u_2) \in E(G_q) \Rightarrow (f(u_1), f(u_2)) \in E(G^{'}_d)$.

Two popular subgraph isomorphism algorithms were developed by Ullman \cite{Ullmann:1976:ASI} and Cordella et al. \cite{cordella2004sub}.  The VF2 algorithm \cite{cordella2004sub} employs a filtering and verification strategy and outperforms the original algorithm by Ullman.  Over the past decade, the database community has focused strongly on developing indexing and query optimization techniques to speed up the searching process.  A common theme of such approaches is to index vertices based on k-hop neighborhood signatures derived from labels and other properties such as degrees, spectral properties and centrality  \cite{tian2008tale, Tong:2007:FBP:1281192.1281271, Zhao:2010:GQO:1920841.1920887, Zhu:2011:HEQ:2063576.2063831}.   Other major areas of work involve join-order optimization \cite{ Zou:2009:DPM:1687627.1687727} and search techniques for alternative representations such as similarity search in a multi-dimensional vector space \cite{Khan:2011:NBF:1989323.1989418}. 


\section{Related Work}
\label{sec:Related Work}
Investigation of subgraph isomorphism for dynamic graphs did not receive much attention until recently.  It introduces new algorithmic challenges because we can-not afford to index a dynamic graph frequently enough for applications with real-time constraints.  In fact this is a problem with searches on large static graphs as well \cite{DBLP:journals/pvldb/SunWWSL12}.  There are two alternatives in that direction.  We can search for a pattern repeatedly or we can adopt an incremental approach.  The work by Fan et al. \cite{Fan:2011:IGP:1989323.1989420} presents incremental algorithms for graph pattern matching.  However, their solution to subgraph isomorphism is based on the repeated search strategy.  Chen et al. \cite{Chen:2010:CSP:1850481.1850517} proposed a feature structure called the \textsl{node-neighbor tree} to search multiple graph streams using a vector space approach.  They relax the exact match requirement and require significant pre-processing on the graph stream.  Our work is distinguished by its focus on temporal queries and handling of partial matches as they are tracked over time using a novel data structure.  From a data-organization perspective,  the SJ-Tree approach has similarities with the Closure-Tree \cite{He:2006:CIS:1129754.1129898}.  However, the closure-tree approach assumes a database of independent graphs and the underlying data is not dynamic.  There are strong parallels between our algorithm and the very recent work by Sun et al. \cite{DBLP:journals/pvldb/SunWWSL12}, where they implement a query-decomposition based algorithm for searching a large static graph in a distributed environment.  Our work is distinguished by the focus on continuous queries that involves maintenance of partial matches as driven by the query decomposition structure, and optimizations for real-time query processing.

\section{Incremental Query Processing}
\label{sec:Incremental Query Processing}

\subsection{Naive Approach}
We begin with a simplistic solution to motivate an incremental approach for continuous query processing.  For every new edge that is added to $G_d$, we detect if the edge matches any edge in the query graph.  This check can be performed minimally by examining 1) if there are edges in the query graph with the same type as the new edge and 2) if the endpoint vertices of the new edge match with the corresponding edges in the query graph based on their types and attributes.  Once an edge is considered as a matching candidate, the next step is to consider different combinations of matches it can participate in.

\begin{figure}[]
\centering
\includegraphics[scale=0.6]{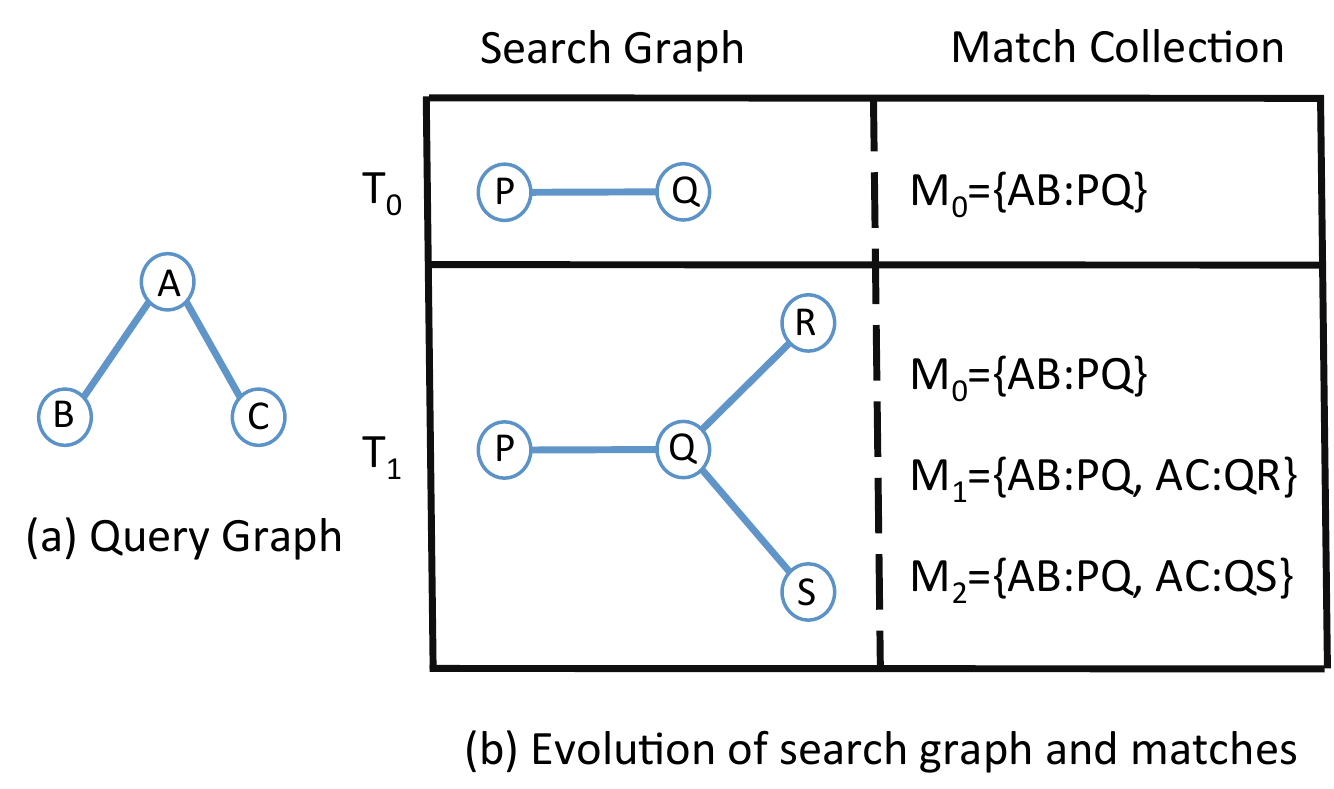}
\caption{Illustration of a naive incremental algorithm.  Assme AB matches with PQ, and AC matches with both QR and QS.}
\label{fig:fig_algo_naive}
\end{figure}

A simple illustration of this matching process is shown in Fig. \ref{fig:fig_algo_naive}.  While intuitively simple, this approach falls prey to combinatorial explosion very quickly.   It finds the match with the query graph at the cost of creating many partial matches.  Assume that the $G_d$ receives a large number of edges that match with the query graph edge between A and B (Fig. 3a).  Let's denote this edge as $e_{AB}$.  Therefore, a large number of partial matches will be created with mapping information for $e_{AB}$.  Subsequently, every future edge that matches with $e_{AC}$ will need to be matched or checked against all the existing partial matches for augmenting into a larger match.  While the subgraph matching problem has an inherent exponential nature associated with it, a better algorithm will restrict the growth of the number of partial matches to track and still produce the correct result.    

\subsection{Our Approach}
Our objective is to introduce an approach that guides the search process to look for specific subgraphs of the query graph and follow specific transitions from small to larger matches.  Following are the main intuitions that drive this approach,
\begin{enumerate}
\item Instead of looking for a match with the entire graph or just any edge of the query graph, partition the query graph into smaller subgraphs and search for them.
\item Track the matches with individual subgraphs and combine them to produce progressively larger matches.
\item Define a \textit{join order} in which the individual matching subgraphs will be combined.  Do not look for every possible way to combine the matching subgraphs.
\end{enumerate}

Although the current work is completely focused on temporal queries, the graph decomposition approach is suited for a broader class of applications and queries.  The key aspect here is to search for substructures without incurring too much cost.  Even if some subgraphs of the query graph are matched in the data, we will not attempt to assemble the matches together without following the join order.  Thus, if there are substructures that are too frequent, joining them and producing larger partial matches will be too expensive without a stronger guarantee of finding a complete match.  On the other hand, if there is a substructure in the query that is rare or indicates high selectivity, we should start assembling the partial matches together only after that substructure is matched.  

\begin{figure}[]
\includegraphics[scale=0.42]{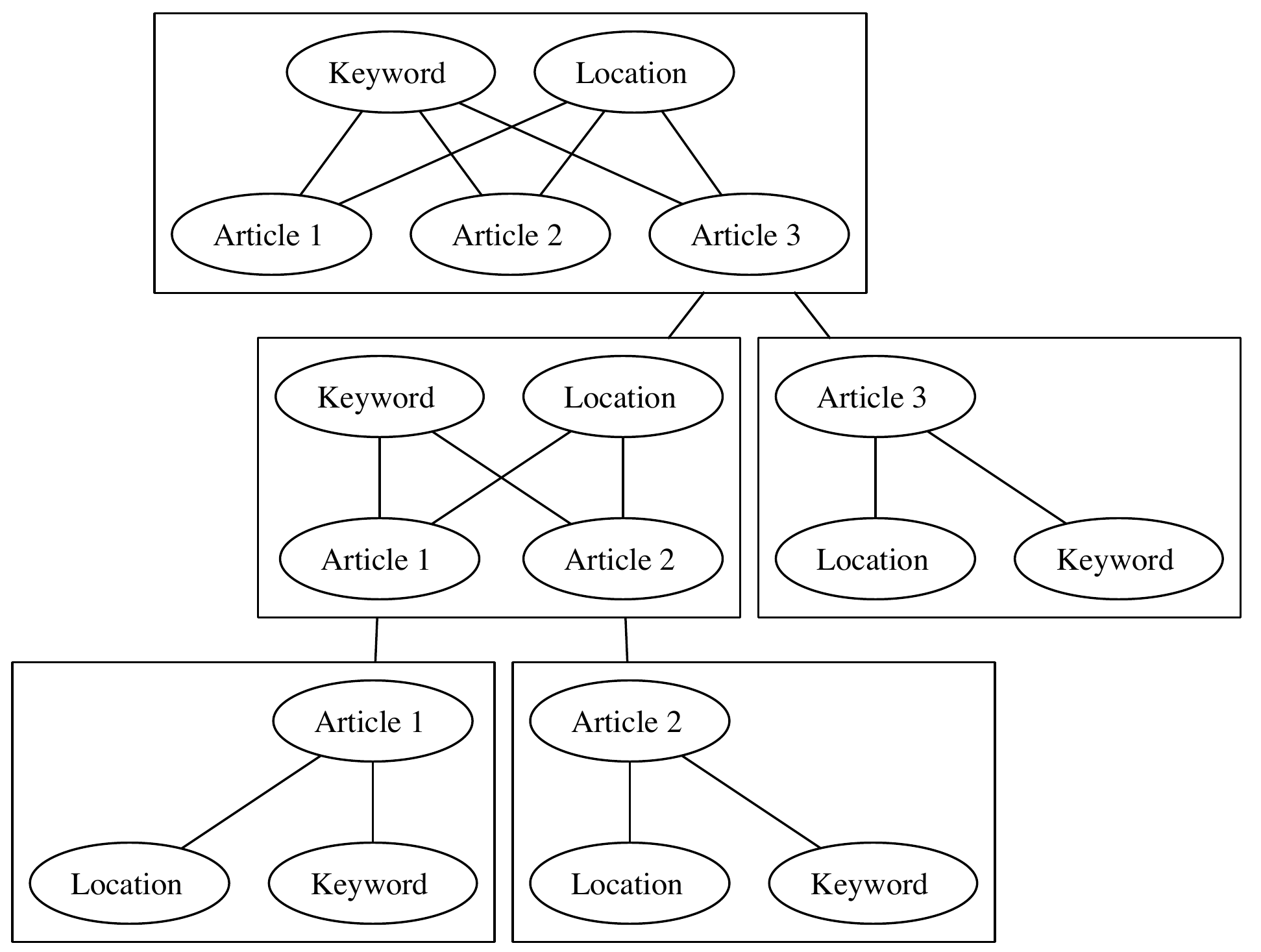}
\caption{Illustration of query decomposition in SJ-Tree. }
\label{fig:join_tree}
\end{figure}

\subsection{Subgraph Join Tree (SJ-Tree)}
\label{subsubsec:join_tree_properties}
We introduce a tree structure called \emph{Subgraph Join Tree (SJ-Tree)} that supports the above intuitions for implementing a join order based on selectivity of substructures of the query graph. 

\textsc{Definition 4.1.1 } A SJ-Tree $T$ is defined as a binary tree comprised of the node set $N_T$.  Each $n \in N_{T}$ corresponds to a subgraph of the query graph $G_q$.   Let's assume $V_{SG}$ is the set of corresponding subgraphs and $|V_{SG}| = |N_T|$.  Additional properties of the SJ-Tree are defined below.


\textsc{Property 1.} The subgraph corresponding to the root of the SJ-Tree is isomorphic to the query graph.  Thus, for $n_r = root\lbrace T\rbrace$, $V_{SG}\lbrace n_r \rbrace \equiv G_q$.

\textsc{Property 2.} The subgraph corresponding to any internal node of $T$ is isomorphic to the output of the join operation between the subgraphs corresponding to its children.  If $n_l$ and $n_r$ are the left and right child of $n$, then $V_{SG}\lbrace n \rbrace = V_{SG}\lbrace n_l \rbrace \Join V_{SG}\lbrace n_r \rbrace$.  Given two graphs $G_1 = (V_1, E_1)$ and $G_2 = (V_2, E_2)$, the join operation is defined as $G_3 = G_1 \Join G_2$, such that $G_3 = (V_3, E_3)$ where $V_3 = V_1 \cup V_2$ and $E_3 = E_1 \cup E_2$.

\textsc{Property 3.} Each node in the SJ-Tree maintains a set of matching subgraphs.  We define a function $matches(n)$ that for any node $n \in N_T$, returns a set of subgraphs of the data graph.  If $M = matches(n)$, then $\forall G_m \in M$,  $G_m \equiv V_{SG}\lbrace n \rbrace$.

\textsc{Property 4.} Each internal node $n$ in the SJ-Tree maintains a subgraph, CUT-SUBGRAPH($n$) that 
equals the \textit{intersection} of the query subgraphs of its child nodes.

 
For any internal node $n \in N_T$ such that CUT-SUBGRAPH$(n) \neq \emptyset$, we also define a \textit{projection operator} $\Pi$.  Assume that $G_1$ and $G_2$ are isomorphic, $G_1 \equiv G_2$.  Also define $\Phi_V$ and $\Phi_E$ as functions that define the bijective mapping between the vertices and edges of $G_1$ and $G_2$.  Consider $g_1$, a subgraph of $G_1$: $g_1 \subseteq G_1$.  Then $g_2 = \Pi(G_2, g_1)$ is a subgraph of $G_2$ such that $V(g_2) = \Phi_V(V\left(g_1\right))$ and $E(g_2) = \Phi_E(E\left(g_1\right))$.
 

Conceptually, the SJ-Tree is an index structure where keys track the occurrence of matching subgraphs in the data graph.   Our decision to use a binary tree as opposed to an n-ary tree is influenced by the simplicity and lowering the combinatorial cost of joining matches from multiple children.  Analyzing the trade-offs between different tree models (e.g. left-deep vs bushy) is part of future work.

\section{Continuous Query Algorithm}
\label{sec:Continuous Query Algorithm}

\begin{figure*}[]
\includegraphics[scale=0.6]{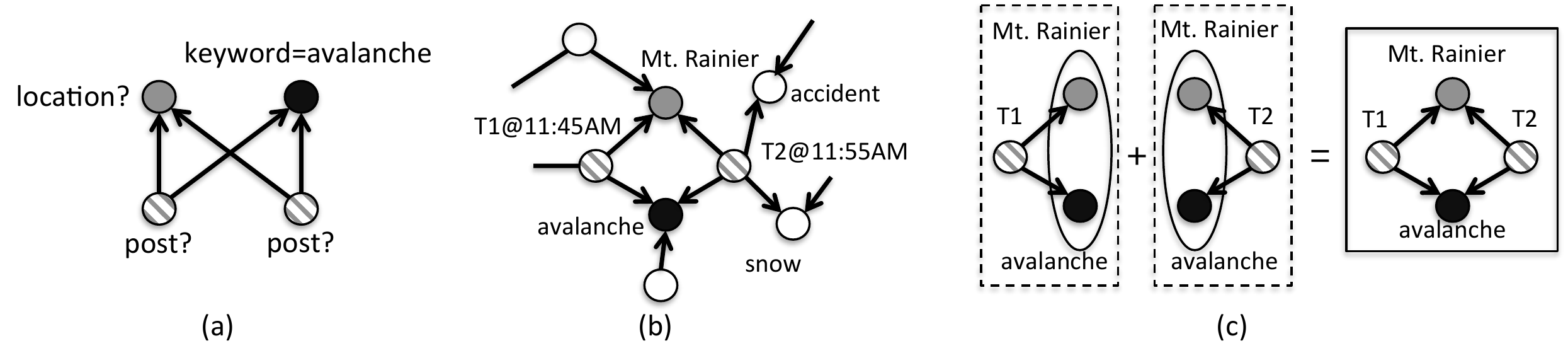}
\caption{a) Example query b) Occurrence of a match in the search graph c) Combining two partial matches to form the complete match}
\label{fig:match_join}
\end{figure*}

We present a subgraph search algorithm (Algo. \ref{algo:process_cont_query} and \ref{algo:update_match_tree}) that utilizes the SJ-Tree structure (referred to as $T$).  The search process is illustrated in Fig. \ref{fig:match_join}.  The input to PROCESS-CONT-QUERY is the dynamic graph $G_d$, the SJ-Tree ($T$) corresponding to the query graph and the set of incoming edges.  Each leaf of the SJ-Tree represents an unique subgraph of the query graph.  Lines 4-6 in Algo. \ref{algo:process_cont_query} describe the search for each of these unique subgraphs around every incoming edge.  Any discovered match is added to the SJ-Tree (line 9).

\begin{algorithm}
\caption{PROCESS-CONT-QUERY($G_d$, T, edges)}
\label{algo:process_cont_query}
\begin{algorithmic}[1]
\State $leaf$-$nodes = $GET-LEAF-NODES$(T)$
\ForAll {$e_s \in edges$}
	\State UPDATE-GRAPH($G_d, e_s$)
	\ForAll {$n \in leaf$-$nodes$}
		\State $g^q_{sub} = $GET-QUERY-SUBGRAPH$(T, n)$
		\State $matches = $LOCAL-SEARCH($G_d, g^q_{sub}, e$)
		\If {$matches \neq \emptyset$}
			\ForAll {$m \in matches$}
				\State T.UPDATE-SJ-TREE$(n, m)$
			\EndFor
		\EndIf
\EndFor
\EndFor
\end{algorithmic}
\end{algorithm}
\subsection{Local Search}

The LOCAL-SEARCH function performs a subgraph isomorphism check around the neighborhood of every incoming edge $e$.  The query decomposition often reduces the local search to performing star queries where the center of the query is the vertex representing a temporal event.  The peripheral vertices of the star query are the other entities that represent various attributes of the event.  Further, in the context of real-time search, if the current time is $t$ and the query specifies a time window of length $t_W$ then all edges that have a timestamp older than $(t - t_W)$ are ignored from the search.  In addition to filtering search candidates, we also periodically prune the SJ-Tree to remove partial matches that are older than $t_W$ from the current time.


\begin{algorithm}
\caption{UPDATE-SJ-TREE($node, m)$}
\label{algo:update_match_tree}
\begin{algorithmic}[1]
\State $sibling = sibling[node]$
\State $parent = parent[node]$
\State $k = $GET-JOIN-KEY(CUT-SUBGRAPH[$parent$], $m$)
\State $H_s$ = match-tables[$sibling$]
\State $M^k_s$ = GET($H_s, k$)
\ForAll {$m_s \in M^k_s$}
\State $m_{sup}$ = JOIN($m_s m$)
		\If {parent = root}
				\State PRINT('MATCH FOUND :  ', $m_{sup}$)				
		\Else
				\State UPDATE-SJ-TREE($parent, m_{sup}$)
		\EndIf
\EndFor
\State ADD($matches[node], m$)
\State ADD($match-tables[node], k, m$)
\end{algorithmic}
\end{algorithm}

\subsection{Partial Match Join and Aggregation}
\label{subsec:temporal_opt}
This subsection describes the process outlined in UPDATE-SJ-TREE.  The SJ-Tree data structure maintains the sibling and parent information for every node as distinct arrays (Algo. 2, line 1-2).  Each node in the SJ-Tree maintains a hash table that supports storing multiple objects with the same key.  This collection of tables are denoted by the match-tables property of the SJ-Tree (Algo 2., line 4).  GET() and ADD() provides lookup and update operations on the hash tables.  Whenever a new matching subgraph $g$ (available as a property of the partial match $m$) is added to a node in the SJ-Tree, we compute a key using its projection $(\Pi(g))$ and insert the key and the matching subgraph into the hash table.  The projection is obtained by hashing a string representation of the subgraph.  

When a new match is inserted into a leaf node we check to see if it can be combined (referred as JOIN()) with any matches that are contained in the collection maintained at its sibling node.  A successful combination of matching subgraphs between the leaf and its sibling node leads to the insertion of a larger match at the parent node.  This process is repeated as long as larger matching subgraphs can be produced by moving up in the SJ-Tree.  A complete match is found when two matches belonging to the children of the root node are combined successfully.

The JOIN operation between partial matches is critical to the overall query processing performance.  Suppose we have a query that finds a sequence of two events with a common set of attributes.  Assume that two matching events ${event}_1$ and ${event}_2$ are found with timestamps $\tau_1$ and $\tau_2$ respectively, with $\tau_1 < \tau_2$.  For all practical purposes we can report the sequence $\lbrace {event}_1, {event}_2 \rbrace$ and ignore the out of order combination.  Therefore, given two partial matches $M_1$ and $M_2$ with edge sets $\lbrace E_i, E_j \rbrace$ and $\lbrace E_m, E_n \rbrace$ respectively, the join algorithm rejects all combinations of these two sets that do not represent a monotonic order based on timestamps.  This is accomplished by computing a range of timestamps for each partial match.  If $t_{low}[M_i]$ and $t_{high}[M_i]$ are the lowest and highest timestamp for match $M_i$, then we require that $t_{high}[M_1] < t_{low}[M_2]$ for joining $M_1$ and $M_2$.  

\subsection{Complexity Analysis}

There are two primary tasks in processing every edge in the continuous query algorithm, (1) performing a local search for a small subgraph of the query graph and in case of a successful search, (2) updating the SJ-Tree with the partial match.  For the multi-relational queries described in this paper, the local search reduces to performing a star query.  Assuming the LOCAL-SEARCH is cheap for star queries, we can approximate the cost of the continuous query processing for every edge to a small constant in case of a failed local search and to that of the UPDATE-SJ-TREE() for a successful search.  If $C_k$ is the expected value for the cost of insert and joins at node $k$ in the SJ-Tree, then the complexity of updating the tree is $O((\bar{C_k})^h)$, where $\bar{C_k}$ denotes the expected value of $C_k$ over all nodes.   This is not a tight bound, and a more accurate bound can be obtained estimating $C_k$ based on the frequency of the subgraphs satisfying the label constraints in the query graph.  Accurate estimation of the frequency of an arbitrary subgraph is hard.  Therefore, we resort to obtaining a loose bound in terms of the label constraints. Assume that $v_q$ has the lowest degree among all labeled vertices in the query graph and $v_s$ is the corresponding vertex in the data graph.  Then $n_k $ is bounded by ${d_s}\choose{d_q}$, where $d_s$ and $d_q$ are the degrees of $v_s$ and $v_q$.  The storage complexity for SJ-Tree is $O(h\bar{C_k} |E(G_q)|)$.

\section{Experimental Results}
\label{sec:Experimental Results}

This section is dedicated to answering two questions: 1) How does our continuous graph query algorithm compare with the state of the art?  To answer this, we compare our algorithm's performance with the IncIsoMatch algorithm presented in \cite{Fan:2011:IGP:1989323.1989420}.  2) How does our query algorithm perform on real-world datasets?  We provide the answers from exhaustive experimentation on three real-world datasets through systematic query selection.  \\

\begin{tabular}{lrrcc}
\hline
\hline
Graph dataset & vertices & edges & vertex types & edge types\\
\hline
New York Times  & 39,523 & 68,682 & 4 & 4\\
DBLP  & 3.158M  & 3.26M & 2 & 1\\
Tencent Weibo  & 2.5M  & 89.6M & 4 & 5\\
\hline
\end{tabular}
\\
Our metric is the time to process increments of a fixed number of edges (1k or 100k) whichever is closer to 1\% of the test dataset size.  The times reported only include the time spent in the query processing section.  We use the query template as shown in Fig. \ref{fig:query_template}.  To develop a performance model in terms of the label distribution, we sample the degree distribution of every vertex type and divide the range of the degree distributions into ten intervals.  For each interval, five closest candidate vertices are selected for testing purposes.  Selection of multiple vertices around each bin allows us to systematically observe the impact of increasing the degree of the labeled vertex in the query graph.  

\subsection{Experimental Setup}
The results were obtained by using a single core on a 48-core shared memory system comprising 2.3 GHz Opteron 6176 SE processors and 256 GB RAM.  The processor cache size is 512KB and each system node has 32 GB RAM.  The code was compiled with g++ 4.1.2-52 with -O3 optimization flag on Linux 2.6.18.

\subsection{New York Times}

\begin{figure}[]
\centering
\includegraphics[scale=0.4]{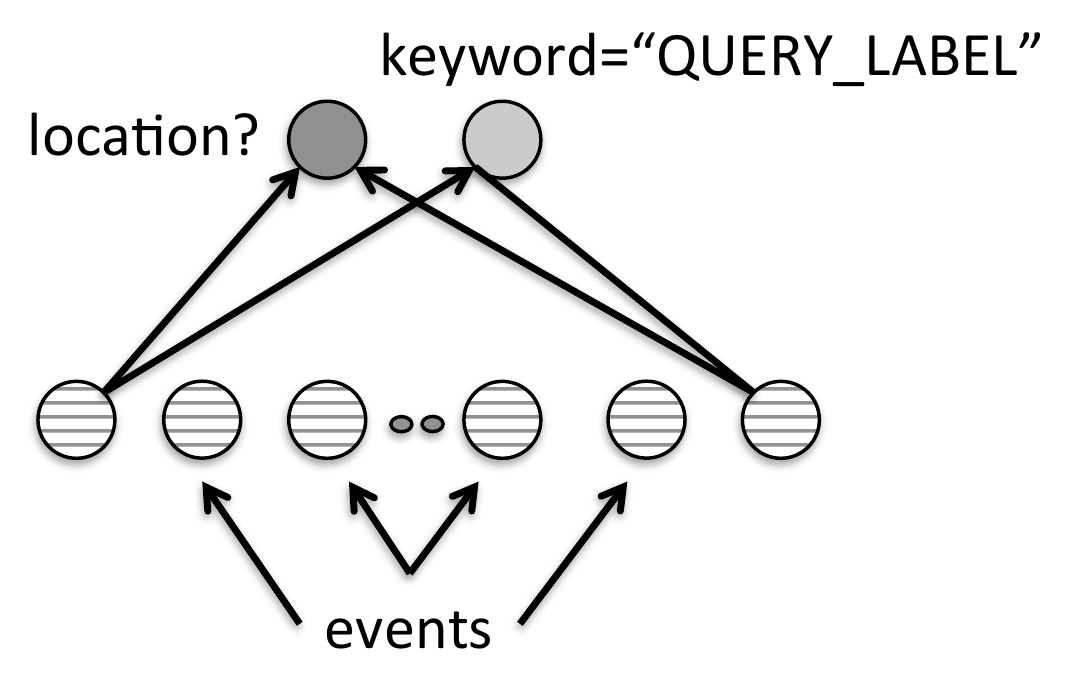}
\caption{The query template used to find temporal events.  Experiments are performed using queries with 4 event vertices and 2 feature vertices.  One feature vertex is labeled and all other vertices specify only types.}
\label{fig:query_template}
\end{figure}
\begin{figure}[]
\centering
\includegraphics[scale=0.4]{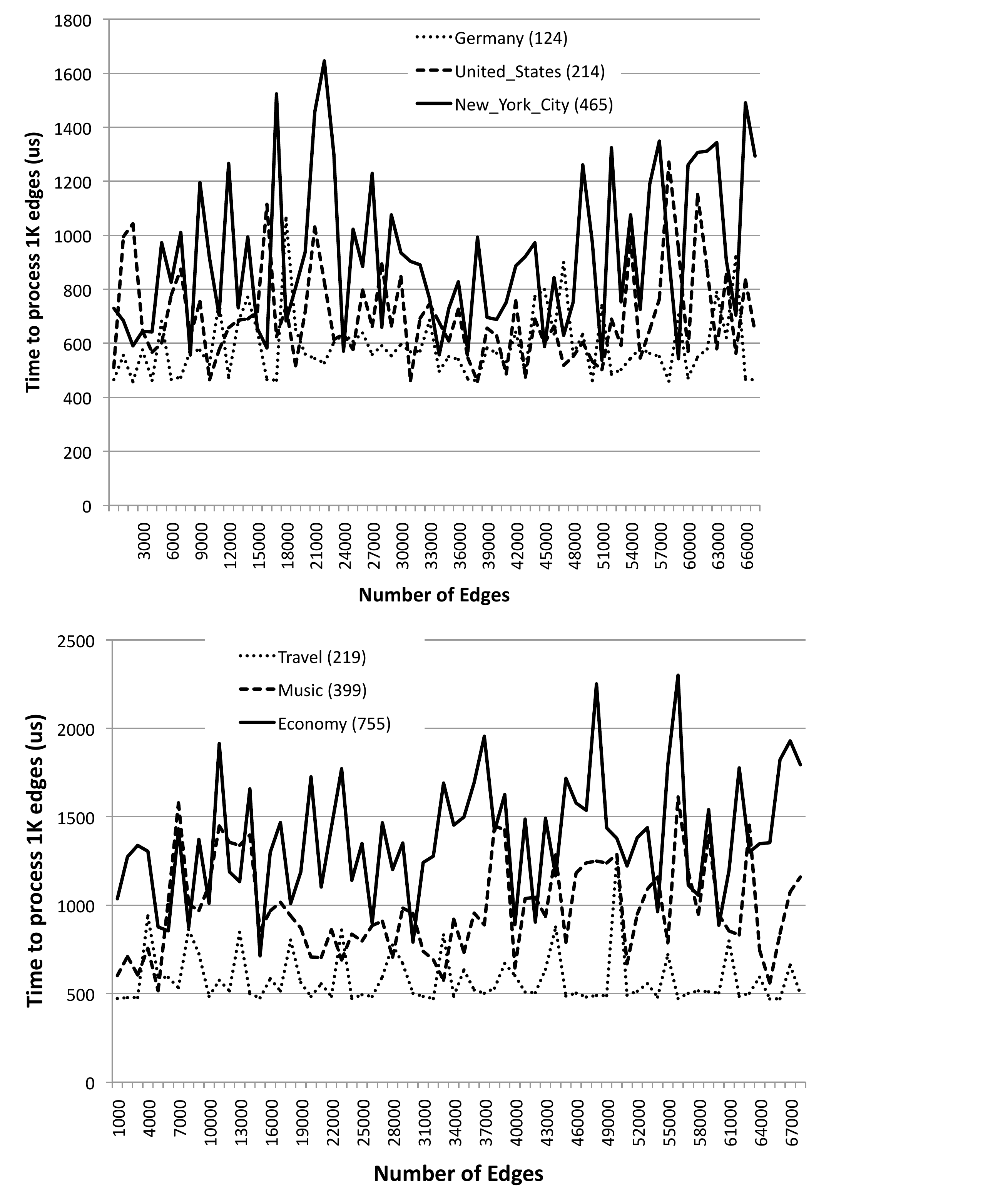}
\caption{Results from queries finding four articles with a common keyword and location.  Legends indicate degree of query label.}
\label{fig:nyt_T4}
\end{figure}
\begin{figure}[]
\centering
\includegraphics[scale=0.3]{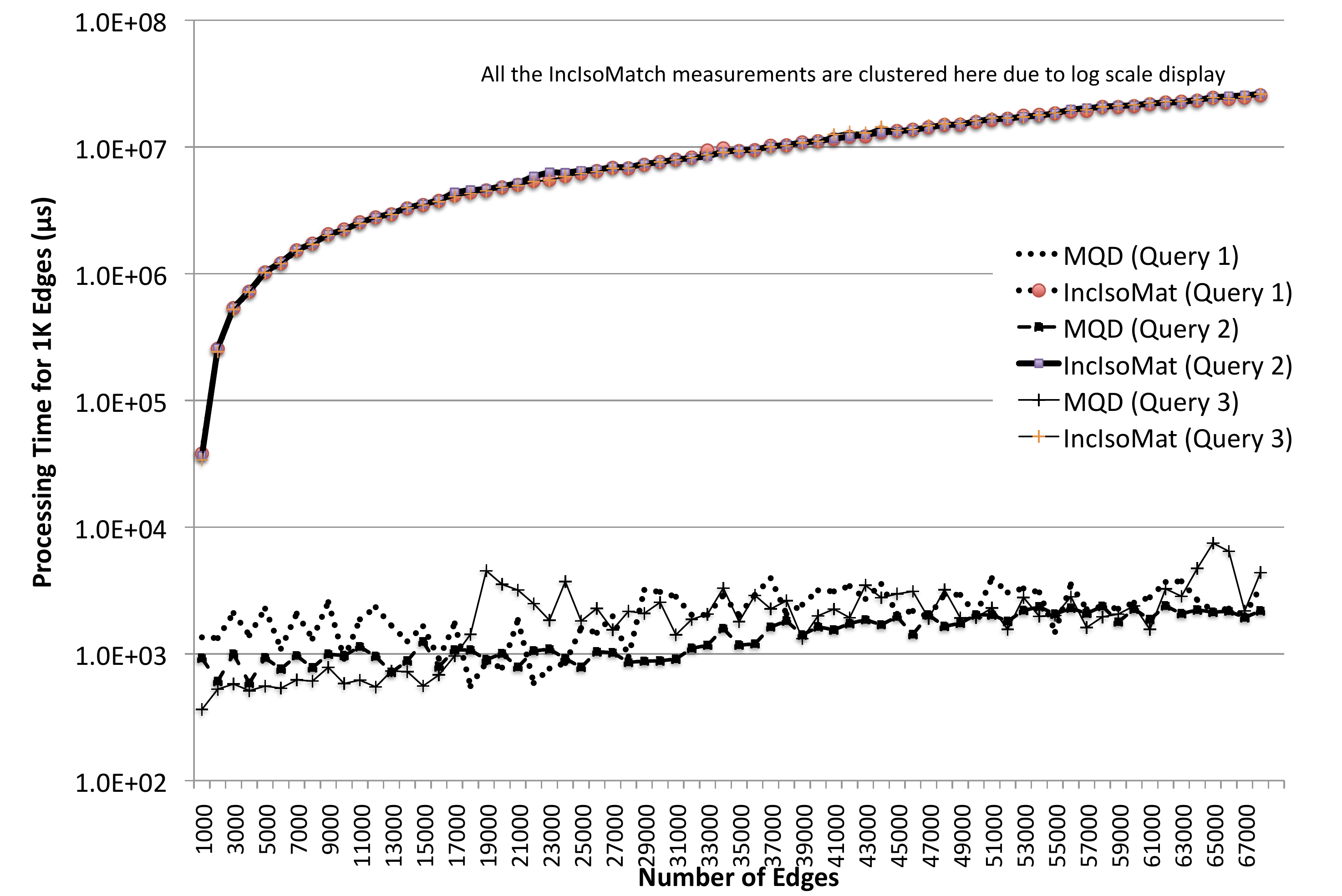}
\caption{Comparison with IncIsoMatch.  MQD refers to the Multi-Relational Query Decomposition algorithm from this paper.}
\label{fig:compare_fan}
\end{figure}


We use a news dataset from New York Times collected over Aug-Oct 2011 \footnote[1]{http://data.nytimes.com}.  Each article in the dataset contains a number of facets that belong to four type of entities.  Each of the articles and facets are represented as vertices in the graph.  Each edge that connects an article with a facet carries a timestamp that is the publication time of the article.  Following the template shown in Fig. \ref{fig:query_template}, we run a query that finds four articles where all the articles have a common keyword and location.  For the location vertex we specify the labels shown in the top diagram in Fig. \ref{fig:nyt_T4} and observe the performance.  As the figures indicate, selecting labels that correspond to vertices with increasing degrees increases the running time of the query.

Next, we compare our approach with the IncIsoMatch algorithm described by Fan et al. \cite{Fan:2011:IGP:1989323.1989420}.  The VF2 algorithm \cite{cordella2004sub} is adapted to implement the graph search functionality as outlined in IncIsoMatch.   We specify a label on the feature marked with $\dagger$ and select a label with one of the highest degrees for that vertex type.  The queries are as follows: 1) Find four articles with a common keyword and a common organization$\dagger$, 2) Find four articles with a common entity  and a common keyword $\dagger$ and 3) Find four articles with a common entity and a common location $\dagger$.  Fig. \ref{fig:compare_fan} shows a performance improvement from our algorithm by several orders of magnitude.  The multiple orders of improvement in performance is attributed to the strictly ordered aggregation of partial matches in the SJ-Tree and the temporal property based optimizations.  The performance gap between the processing time of the two algorithms increases as the graph grows larger.  We attribute this to the nature of the IncIsoMatch where it performs a search around every new edge in the graph.  The search spans all vertices around the endpoints of the new edge as long as they are within $k$-hops, where $k$ is the diameter of the query graph.  As the data graph grows denser, even for a query graph with small or modest size, the $k$-hop subgraph accumulates a large number of edges and the search becomes increasingly expensive.

\subsection{DBLP Co-Authorship Network}
\begin{figure}[]
\centering
\includegraphics[scale=0.38]{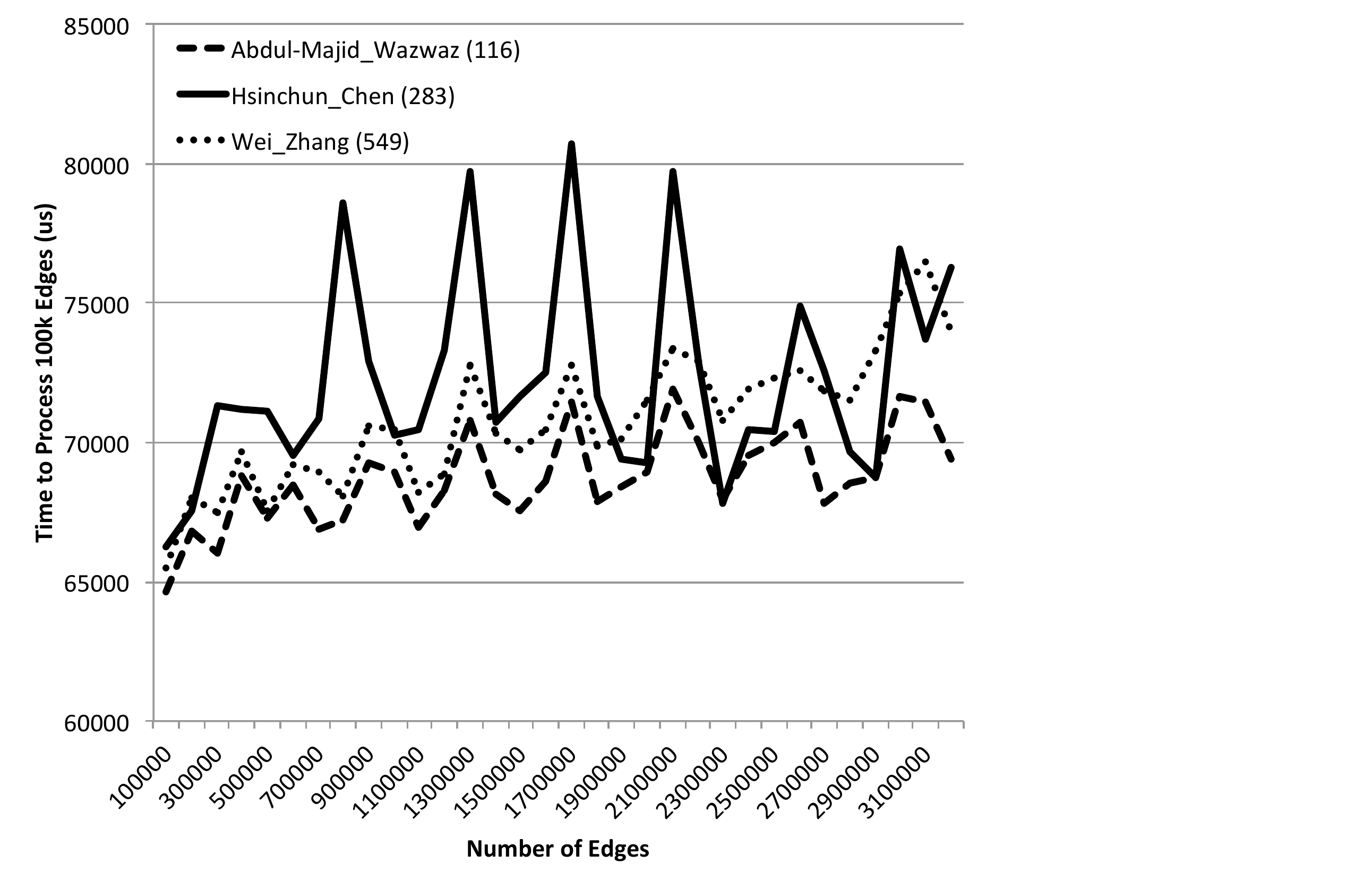}
\caption{Performance results for queries on the DBLP dataset.  The spikes in the plot can be attributed to the bursty nature of scientific publishing where authors target the same group of conferences and journals every year.}
\label{fig:dblp_query_perf}
\end{figure}
We build a multi-relational graph representation of the DBLP citation network \footnote[1]{dblp.uni-trier.de/xml} with two types of entities: authors and articles.  The author name and the title of the article are stored as labels of respective vertices.  We run a query to find an author (author 1) who has co-authored four papers with a specified author (author 2).  Following the previously shown query template, our query graph has four article vertices and two author vertices.  Only one author vertex is labeled.  We observe the degree distribution of the ``author" vertices and select names with progressively increasing degrees.  The results are shown in Fig. \ref{fig:dblp_query_perf}.  It can be seen that the performance of the algorithm is quite stable for a modestly large network with nearly 3M+ edges.  Additionally, the results show that even though vertex degree is a good indicator of the query performance, there are other factors at play.  The graph describes author-article relationship; therefore, the degree of an author vertex provides the number of authored articles.  It does not provide the information about the number of co-authors of a person.  Searching for a person who publishes a given number of articles with fewer co-authors will lead to more partial matches and increase per-edge query processing time.  The consistent high processing times for the query containing "Hsinchun-Chen" despite smaller degree in the graph is a result of this aspect.

\begin{figure}[]
\centering
\includegraphics[scale=0.35]{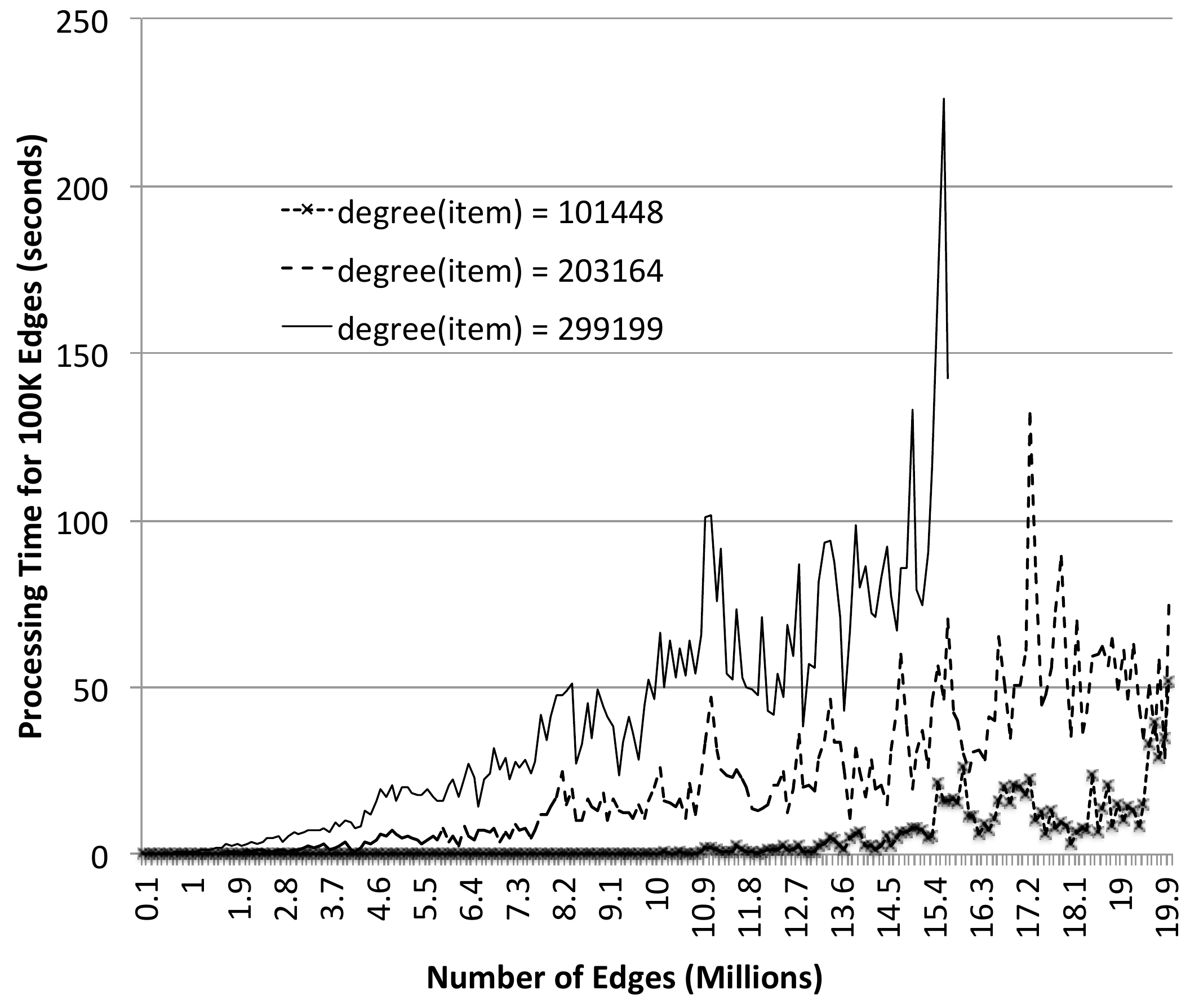}
\caption{Query processing time for the Tencent Weibo dataset for queries with varying selectivity.}
\label{fig:kddcup12_degree_plot}
\end{figure}

\subsection{Social Media}

Finally, we present our results on a data set collected from Tencent Weibo, a Chinese microblogging social network\footnote[2]{www.kddcup2012.org/c/kddcup2012-track1}.  The data set provides a temporal history of item recommendations to registered users of the social network. We build a graph with 4 vertex types (users, items, keywords and categories) and 5 edge types (item-in-category, item-has-keyword, item-reco-accept, item-reco-reject and user-profile-has-keyword).  

Our test query is to detect a series of item acceptances by a group of users described by a common keyword.  Following the previously shown query template, our query graph has four user vertices and one item and keyword vertex.  We specify a label on the item and seek to discover the keyword that characterize the users accepting or rejecting that item.  The results are shown in Fig. \ref{fig:kddcup12_degree_plot}.  The figure suggests a clear trend.  It shows that as the graph grows large the query processing time eventually rises sharply.   It also shows that the rise happens earlier for low-selectivity queries where the specified label has higher degree in the graph.  This is because the number of partial matches grows rapidly in the event of a successful search around a high degree vertex.  Every partial match from the past can potentially be merged with the latest partial match, and the partial match collection grows combinatorially over time.  

\begin{figure}[]
\centering
\includegraphics[scale=0.35]{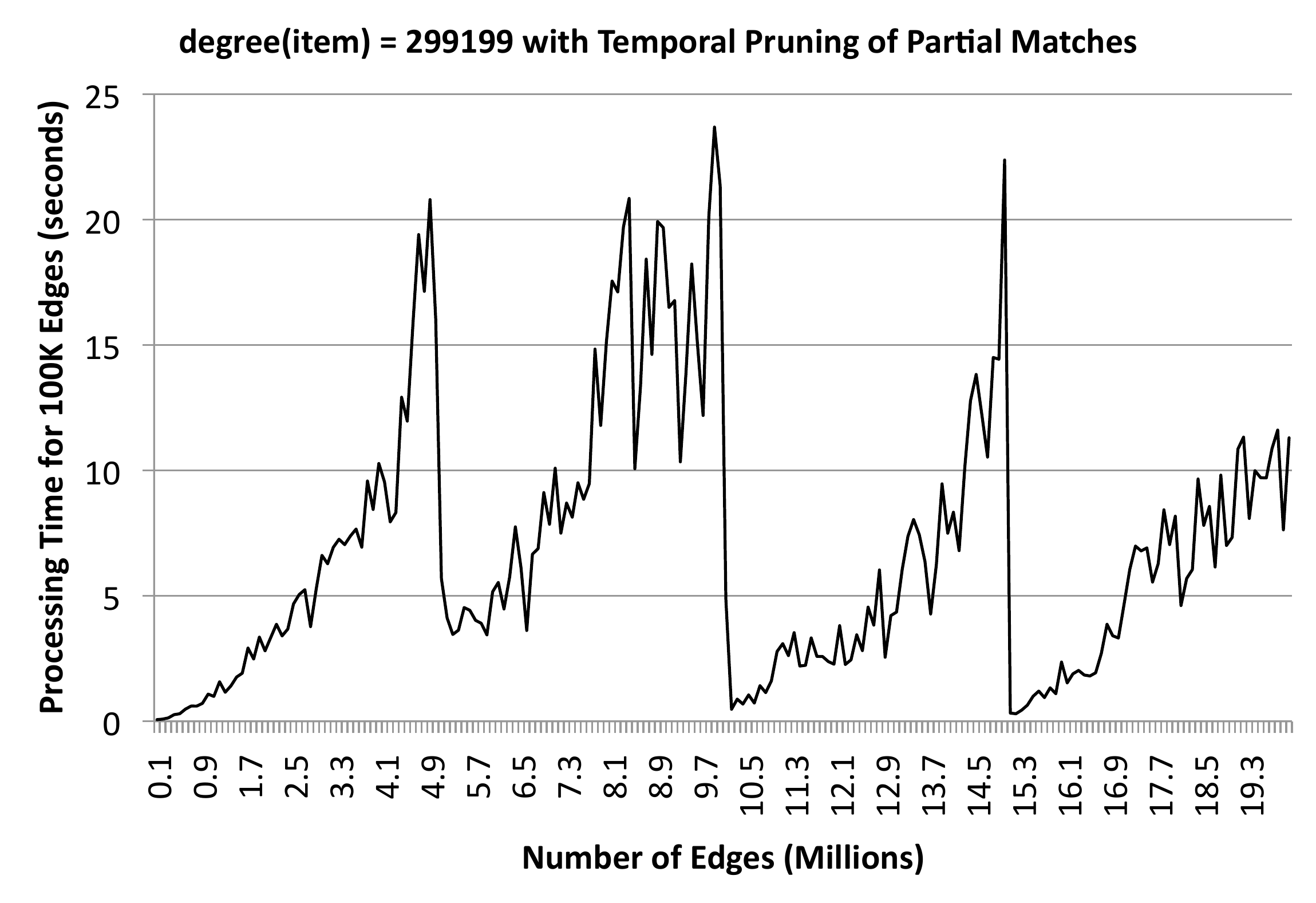}
\caption{Query processing time for the Tencent Weibo dataset with temporal match pruning applied on every 5 million edges.}
\label{fig:kddcup12_pruning}
\end{figure}

This brings us to implementing the temporal window based pruning.  We select the query with the highest degree label (degree(item) = 299199, Fig. \ref{fig:kddcup12_degree_plot}) for which the rise in the processing time was sharpest.  We set the time window $t_W$ to 1 day and prune the SJ-Tree after processing every 5 million edges.  The results from the windowing enabled search is shown in Fig. \ref{fig:kddcup12_pruning}.  Observe that the peaks in the processing time are smaller than ones observed in Fig. \ref{fig:kddcup12_degree_plot} by an order of magnitude.

This is an extremely promising result for practical applications.  Figure \ref{fig:kddcup12_pruning} suggests that it would take 10 seconds on average to process 100k edges for a query with very low selectivity.  This translates into a throughput of 0.01 million edges/second or 864 million edges per day.  At the time of this writing, high volume data streams such as Twitter receive nearly 300-400 million posts every day.  Considering that every user action translates into multiple edges in a graph, one may expect around billions of edges everyday.  The throughput can be expected to be much higher for a query with moderate selectivity.   Thus, we believe this level of throughput on a very low-selectivity query gets us close to executing real-time graph queries on such high volume data streams.

\section{Conclusion and Future Work}
\label{sec:Conclusion and Future Work}

We present a novel graph decomposition based approach for continuous queries on multi-relational graphs.  We introduce the SJ-Tree structure, whose nodes represent the hierarchical decomposition of the query graph.  The SJ-Tree systematically tracks the evolving matches in the data graph as they transition from smaller to larger matches based on the query graph decomposition.  We present experimental analysis on several real-world datasets such as New York Times, DBLP and Tencent Weibo and show that our SJ-Tree based algorithm coupled with temporal optimizations clearly outperforms the state of the art \cite{Fan:2011:IGP:1989323.1989420} by multiple orders of magnitude.  Our experiments demonstrate that it is possible to execute complex multi-relational graph queries in a real-time setting.  To our knowledge, the results presented in this paper are the best reported performance for such queries.  These initial results are highly promising in that they suggest possible ways of auto-selecting optimal values for query processing parameters based on the data distribution.  Our main theoretical contribution is to demonstrate that for a prominent class of multi-relational queries where the \textsl{local search} is cheap, we can execute graph queries in time that is exponential to the height of the SJ-Tree.  

Development of query planning algorithms to generate a SJ-Tree for any query graph by exploiting its structural and semantic characteristics is the next logical step.  Query planning for 1) complex graph queries where a complete temporal ordering may not be possible, 2) trade-offs between different query decomposition strategies and 3) exploring different query classes and determining the optimal trade-off between local search and joins in the SJ-Tree represent areas of future work.  


\section*{Acknowledgment}
The authors are grateful to Dr. Bill Howe at the University of Washington for his suggestions that have improved this paper.  Presented research is based on work funded under the Center for Adaptive Supercomputing Software - Multithreaded Architecture (CASS-MT) at the US Department of Energy's Pacific Northwest National Laboratory, which is operated by Battelle Memorial Institute.

\bibliographystyle{IEEEtran}
\bibliography{citations_sutanay}

\balancecolumns
\end{document}